# Raman scattering study of the a-GeTe structure and possible mechanism for the amorphous-to-crystal transition

K. S. Andrikopoulos,[1,2] S. N. Yannopoulos,[1,]* G. A. Voyiatzis,[1] A. V. Kolobov,[3,4] M. Ribes,[4] and J. Tominaga[3]

[1] *Foundation for Research and Technology Hellas – Institute of Chemical Engineering and High Temperature Chemical Processes (FORTH / ICE-HT), P.O. Box 1414, GR-26504 Patras, Greece*

[2] *Physics Division, School of Technology, Aristotle University of Thessaloniki, GR-54124, Thessaloniki, Greece*

[3] *Center for Applied Near-Field Optics Research, National Institute of Advanced Industrial Science and Technology, Tsukuba Central 4, 1-1-1 Higashi, Tsukuba, Ibaraki 305-8562, Japan*

[4] *Laboratoire de Physicochimie de la Matiere Condensee, UMR CNRS 5617, Universite Montpellier II, Place Eugene Bataillon, 34095, Montpellier, Cedex 5, France*

**Abstract**

We report on an inelastic (Raman) light scattering study of the local structure of amorphous GeTe films. A detailed analysis of the temperature-reduced Raman spectra has shown that appreciable structural changes occur as a function of temperature. These changes involve modifications of atomic arrangements such as to facilitate the rapid amorphous-to-crystal transformation, which is the major advantage of phase-change materials used in optical data storage media. A particular structural model, supported by polarization analysis, is proposed being compatible with the experimental data as regards both the structure of a-GeTe and the crystallization transition. The remarkable difference between the Raman spectrum of the crystal and the glass can thus naturally be accounted for.

_______________________________
* Corresponding author. E-mail: sny@iceht.forth.gr

1. **Introduction**

GeTe is a very interesting material for both technological applications and fundamental studies. It is well known that GeTe is the basic ingredient of a class of materials (GeTe-$Sb_2Te_3$ quasi-binary alloys) employed as the active medium in modern optical data storage devices, exploiting the phase–change effect [1]. This effect pertains to the reversible amorphous-crystal-melt-amorphous cycle under the action of a laser beam. The particular features of the amorphous-to-crystal transition (erasure process in storage devices) in the ternary Ge-Sb-Te media have only very recently become comprehensible after intense studies of local structure [2].

As far as fundamental studies are concerned, GeTe is famous for two reasons: (i) Although the detailed atomic configuration of the amorphous phase is still in question, old and recent local structure information support that in the amorphous phase not only are long range correlations absent, as expected, but also the short-range structural order is very different from that of the crystal [3, 4]. This renders the understanding of the amorphous-to-crystal transition a challenge, as evident from the numerous recent refined experimental approaches. (ii) Crystalline GeTe is a diatomic ferroelectric material with a ferroelectric-to-paraelectric transition temperature lying in the interval 650-710 K, depending on various factors, such as exact composition, size effects etc. The structural details of this transition are still unknown due to the inherent difficulties of the high temperature studies of this system with the aid of light and neutron scattering. In particular, neutron scattering techniques require massive samples of the order of few grams; making a-GeTe films of such quantity is practically impossible and hence the study of the amorphous-to-crystal transition via neutron scattering is in essence unachievable. Light (Raman) scattering is also difficult to be performed on this specific material at high temperature due to its highly absorbing nature, which renders the scattered signal extremely weak. Increasing laser power provides no solution to the problem since it induces film crystallization. The above demonstrate that there are indeed particular difficulties in studying GeTe at high temperature either with neutron or light scattering.

It is worth-noting that despite the material's importance, e.g. in optical data storage, no temperature dependent Raman scattering study exists to-date across the amorphous-to-crystal and the ferroelectric phase transition. The present paper attempts a study of the first of the two aforementioned structural transitions of GeTe films with the aid of inelastic (Raman) scattering. This is a powerful tool to study transitions where changes on a short length scale take place and hence crystallization-induced short-range structural order changes can be deciphered. The usefulness of Raman scattering in elucidating structural aspects of amorphous semiconductors has long ago been established [5].

The paper is organized as follows. Section 2 presents a brief survey on the structural aspects of amorphous and crystalline GeTe, particularly on the short-range level of structural organization. Section 3 contains the experimental details, namely sample preparation and light scattering apparatus. The results and their discussion alongside with existing structural models are presented in Sec. 4 which is divided in two subsections. In Sec. 4.1 we deal with the structure of amorphous GeTe (a-GeTe) while Sec. 4.2 contains the details of the amorphous-to-crystal phase transition. Finally, the most important conclusions drawn from the present study are summarized in Sec. 5.

## 2. Structural considerations of a-GeTe and c-GeTe

To facilitate the discussion of the results obtained in the present work, as well as to make this paper self-contained, we present in this section a brief survey on some important physical and structural information concerning the amorphous and crystalline forms of GeTe.

Crystalline germanium telluride, c-GeTe, is a semiconductor with a very small bandgap, 0.1-0.2 eV as deduced from transport data [4(b)], whilst optical data result in the value of ~0.8 eV [4(b)]. c-GeTe exhibits also a huge electrical conductivity at room temperature ($3 \times 10^3$ $\Omega$ cm$^{-1}$), close to the metallic limit [4(c)]. It is characterized by high reflectivity of the order of 65% at 633 nm [6]. On the other hand, the amorphous form, a-GeTe, is a semiconductor with a bandgap of 0.8 eV and a low electrical conductivity, almost $10^6$ lower than that of the crystalline form [4(b)]. The reflectivity of a-GeTe is of the order of 40% at 633 nm [6]; appreciably lower than the reflectivity of c-GeTe. The thermal conductivities of c-GeTe and a-GeTe at 300 K are 0.012 and 0.005 cal cm$^{-1}$ K$^{-1}$ s$^{-1}$, respectively [4(d)].

The origin of the stabilization of the rhombohedral (distorted rocksalt) structure of c-GeTe has been sought over the last four decades. Many ideas have been proposed; the most realistic among them attribute the distortion either to (i) the crystallization of GeTe as a non-stoichiometric compound Ge$_{0.98}$Te with 2 at. % Ge vacancies, (ii) the high carrier (hole) conductivity, or (iii) the ratio of the atomic masses of Ge and Te. However, critical analysis of experimental data did not lend support to these mechanisms [for details see 4(a)]. Early results on crystallization kinetics have shown that the heat of transformation amorphous→crystal is of the order of 1.0±0.2 kcal mol$^{-1}$ (0.043 eV) and the films are either completely amorphous below $T^{cr}$ or completely crystalline above $T^{cr}$ [4(a)].

The concept of a contrasting short-range structural order (change in the coordination number) between the crystal and the amorphous forms has been appreciated very early in the literature [4(b)] in an effort to comprehend the significant difference between these two forms. Radial distribution functions of a-GeTe films of stoichiometric compositions obtained by an electron diffraction study revealed a change in the coordination number from 4 (tetrahedral) to 6 (octahedral) in the amorphous→crystal transition [7(a)]. At the same time, x-ray diffraction studies supported this contrasting short-range order environment between a-GeTe and c-GeTe [7(b)]. This latter study suggested that a random covalent model where Ge atoms are fourfold coordinated and Te atoms twofold coordinated, 4(Ge):2(Te), describes the short range-order structural correlation of a-GeTe quite satisfactorily. A detailed Ge K-edge extended x-ray absorption fine structure (EXAFS) study by Maeda *et. al.* [8] revealed that the coordination number of Ge is 3.7; a value close to the tetrahedral value. It was suggested that the random covalent model was suitable for describing the local structure of a-GeTe [8]. More recent EXAFS data concerning interatomic distances appeared in Ref. [3, 9].

Raman scattering and IR absorption studies have been undertaken for the amorphous $Ge_xTe_{1-x}$ system by Fisher, Tauc, and Verhelle [10]. Based on the similarity in mode frequencies over a wide range of composition these authors suggested that their data strongly support the random covalent network of Ref. [7(b)]. Therefore, they presented evidence for the $GeTe_4$ tetrahedral units in GeTe. The results of Fisher *et. al.* will be discussed in more detail in Sec. 4.1 alongside with the data of the present work.

An appreciable piece of information has also been obtained concerning the electronic structure of crystalline and amorphous GeTe. Many works have focused on the difference of the electronic properties of a-GeTe and c-GeTe employing x-ray photoelectron spectroscopy, ultraviolet photoelectron spectroscopy and inverse photoemission spectroscopy [11-14]. All of them lend support to the idea of a local structure with tetrahedral local coordination of a-GeTe. In the most recent detailed work, Hosokawa *et. al.* [14] proposed that a randomly bonded 4(Ge):2(Te) coordination described better the experimental inverse photoemission data. Contrary to the aforementioned works, neutron [15] and Mössbauer spectroscopy [16] studies on a-GeTe have been considered to support the threefold coordination of both the Ge and the Te atoms, namely 3(Ge):3(Te).

3. **Experimental details**

GeTe thin films, of about 100 nm thickness, were deposited by radio-frequency magnetron sputtering directly onto silicon substrates. The film composition was examined by x-ray fluorescence spectroscopy; the $Ge_{50}Te_{50}$ concentration was verified with 1% accuracy. A 20 nm thick $ZnS-SiO_2$ layer was deposited by rf-sputtering (on top of the chalcogenide film) in order to simulate the real devices and hence the information obtained from the present study is presumably more relevant to phase-change data storage than those studies employing single crystals. This is particularly true given that the stresses generated in thin films may affect seriously various physical parameters such as transition temperatures. The as–prepared amorphous a-GeTe film was placed onto a hot stage (Linkam Ltd., model: THMS600) with controllable temperature range 80-873 K appropriate for use under the microscope of the micro-Raman setup.

The 763 nm excitation emerging from a Ti:Sapphire laser pumped by an $Ar^+$ ion laser was used to record the Raman spectra. The power density on the sample was kept at low levels in order to avoid undesired irradiation-induced heating; we used a power of less than 0.4 mW on the film focused on a 5-μm radius area. The backscattered light was analyzed by a triple monochromator (Jobin-Yvon T64000) operating at double subtractive mode and was then detected by a CCD cooled at 140 K. The spectral resolution for the recorded Stokes-side Raman spectra was set to ~2 $cm^{-1}$ (this resolution was achieved by usage of 1800 grooves/mm gratings and 250 μm slits) in order to achieve maximum signal quality. The temperature controller of the hot stage was maintaining the temperature fixed during each measurement with stability better than 0.1 $^oC$. Accumulation times of the order of ~30 min were sufficient to achieve a satisfactory signal-to-noise ratio. Calibration of the wavenumber scale in order to take into account possible shifts of the monochromator, as well as of the polarization response of the Raman system, has been performed by measuring the appropriate (polarized) spectra of liquid $CCl_4$.

A simple quantitative estimation of the possible temperature rise of the illuminated volume due to the laser radiation absorption could be obtained by the relation $\Delta T \approx I^{abs}/(2\pi r \kappa)$ where $I^{abs}$ denotes the absorbed light energy by the sample, which is dissipated as heat, $r$ is the radius of the focused laser beam and $\kappa$ is the thermal conductivity of the medium. $I^{abs}$ in our case can be estimated from the difference of the incident light energy and the reflected component. Taking into account the particular experimental details of this work and using the reflectivities and thermal conductivities given in the previous section we obtain for a-GeTe and c-GeTe $\Delta T^a \approx 20$ K and $\Delta T^c \approx 1$ K, respectively. Since these differences do not affect the results discussed below we hereafter refer to the experimentally measured temperatures and not to the corrected ones.

## 4. Results and discussion

Stokes-side Raman spectra of the a-GeTe and c-GeTe films for various temperatures are shown in Fig. 1. The respective temperatures are shown beside the curves. Data collection started with the as-deposited a-GeTe films at 82 K. The temperature was progressively increased until the crystallization of the amorphous films occurred at the temperature interval 410–420 K, in close agreement with the literature values [4(a)] taking into account the film thickness employed in the present work. It should be stressed here that although the symmetry of the rocksalt structure does not allow first order Raman scattering, in the distorted rocksalt structure this constraint is lifted and hence first order Raman spectrum can be observed.

By inspection of Fig. 1 it is easily conceived that the amorphous-to-crystal transition is manifested in the Raman spectra through drastic changes. An obvious modification occurs in the low frequency region (below ca. 80 cm$^{-1}$); compare for example the 300 K spectra for a-GeTe and c-GeTe. The former (a-GeTe) is dominated by the low-energy excitations, namely the Boson peak, and other modes as well, attributed to disorder-induced scattering that results from the breakdown of the conservation of the selections rules in the amorphous medium. However, a detailed discussion of the low-frequency excitations could not be safely advanced here due to the strong interference of the huge elastic peak over the spectral region of interest. We thus concentrate on the high frequency vibrational modes. Before this, we should note the unusual similarity between the amorphous and crystalline phases, which originates from both the wider c-GeTe Raman peaks compared to other crystalline solids and the narrower a-GeTe Raman peaks in comparison to other glasses and amorphous solids. This fact suggests a rather alike degree of topological "disordering" between these two forms. In particular, c-GeTe (having the distorted rocksalt structure) is more disordered than normal crystals while a-GeTe seems more ordered than normal glasses.

An interesting feature of the Raman spectra is the appreciable intensity changes of the vibrational modes with increasing temperature even in a temperature range below $T^{cr}$. This is to be contrasted with the vast majority of amorphous and glassy solids that below their glass transition temperatures exhibit subtle intensity changes of the Raman bands, limited in most of the cases to a moderate peak broadening and negligible red-shift.

In order to proceed to a quantitative determination of the temperature dependence of the amorphous spectra as well as the amorphous-to-crystal transition it is imperative to use a peak fitting procedure with the aim to separate individual vibrational modes from the composite line envelopes of a-GeTe. In doing so, we have to take into account the fact that the relative intensities of the vibrational lines in a Raman spectrum depend – among other factors – on the population of the vibrational energy levels involved in the scattering process, and the latter are amenable to temperature variations. To remove the influence of temperature, and hence to monitor pure structural changes we employ the so-called *reduced representation* [17]. Due to the Boson–like statistical nature of phonons, their mean number at any temperature is given by $n(\omega,T) = [\exp(\hbar\omega/k_B T) - 1]^{-1}$ where $\hbar$ and $k_B$ are the Planck and Boltzmann constants, respectively. Therefore, the Stokes–side reduced Raman intensity ($I^{red}$) is related to the experimentally measured one ($I^{exp}$) through the following equation [18],

$$I^{red}(\omega) = (\omega_0 - \omega)^{-4} \omega [n(\omega,T) + 1]^{-1} I^{exp}(\omega) \qquad (1)$$

where the term in the fourth power is the usual correction for the wavelength dependence of the scattered intensity; $\omega$ is the Raman shift in cm$^{-1}$, and $\omega_0$ denotes the wavenumber of the incident radiation.

*4.1. Structure and vibrational modes of a-GeTe*

Figure 2 illustrates the reduced representation of the Raman spectra of a-GeTe shown in Fig. 1. Let us first discuss the qualitative features of these spectra. Four main bands appear in each spectrum in the frequency range 50–250 cm$^{-1}$. Approximate wavenumber positions of these bands are as follows. Band A: ~83 cm$^{-1}$; band B: ~125 cm$^{-1}$; band C: ~162 cm$^{-1}$; band D: ~218 cm$^{-1}$, (these values are indicative of the 300 K spectrum). Bands A, B and C are certainly combinations of at least two individual peaks each. Before discussing possible structural models of the a-GeTe → c-GeTe transition we should make clear that the peaks quoted above originate solely from the material under study and no contribution from the ZnS-SiO$_2$ layer is involved. This has been tested by comparing Raman spectra obtained by GeTe films with and without the presence of the protective film; the comparison showed indistinguishable spectra.

Should we had no hint from previous studies (see Section 2) concerning the contrasting local structure between the amorphous (4:2) and the crystalline (3:3) phases one could reach (by inspecting Fig. 2) to the impetuous conclusion that peaks A and B of a-GeTe have a direct correspondence with the two peaks of c-GeTe. Then one has to account for the two high frequency modes C and D. A possible origin of those peaks is the second order scattering mechanism but obviously this scenario is incorrect for the following reasons. The intensity of bands C and D is comparable to the intensities of bands A and B; this effect is not at all usual for second order scattering mechanisms that is usually one to two orders of magnitude weaker than one-phonon scattering. Further, the symmetry of bands C and D is different (see below the discussion about the polarization analysis) from the symmetry of bands A and B. Knowing that overtone or combination bands maintain the symmetry of the bands from which they originate, it is highly unlike that bands C and D are manifestations of second order scattering.

Contrary to the above simplistic picture, the comparison between the spectra of amorphous and crystalline GeTe unequivocally demonstrates the contrasting short-range structural order of these two phases. The Raman spectrum of the crystal (peaks at 80 cm$^{-1}$ and 122 cm$^{-1}$) seems to be red-shifted by ~45 cm$^{-1}$ with respect to the spectrum of the amorphous form. This implies that the distorted rocksalt structure of the crystalline ferroelectric phase is not a good structural candidate for describing the structure of the amorphous state. Below, we will present clear evidence that the tetrahedral coordination is compatible with the Raman data.

In order to reveal the nature of the building blocks of a-GeTe it would be instructive to study the composition dependence of the Raman spectra of the binary system $Ge_xTe_{1-x}$ including both the Te-rich and the Ge-rich phases. Such work has been conducted by Fisher *et. al.* [10]. In brief, their findings are summarized as follows. Room temperature Raman spectra were compared for compositions x: 0.20, 0.33, 0.42, 0.50, and 0.67. However, spectra were recorded with the aid of the 514.5 nm laser line at high power levels, which can produce photo- and/or thermo-structural changes on the films. The poor spectral resolution (12 cm$^{-1}$ compared to 2 cm$^{-1}$ of the present work) did not allow the authors of Ref. [10] to observe fine structure in their bands. The low resolution and the choice of the exciting line led to a surprising outcome: Raman spectra in the composition range 0.33–0.50 were found indistinguishable [10]. Despite the above shortcomings, there is a rather good agreement between the peak positions reported by Fisher *et. al.* and those mentioned above. Specifically, for x=0.50, they found four peaks located at 87, 131, 169, and 228 cm$^{-1}$, respectively [10]. In addition, a broad and very diffuse peak reported at 275 cm$^{-1}$ was attributed to Ge-Ge vibrations since it is apparent only in the Ge-rich systems. It is noteworthy that the dominant Raman mode in Ref. [10] is a peak at 169 cm$^{-1}$ while in the present work band B at 125 cm$^{-1}$ is the strongest one in the room temperature Raman spectrum. The strongest vibrational mode of a-Te (symmetric stretching vibration of Te-Te bond) is located at about 150 cm$^{-1}$ while that of crystalline (trigonal) c-Te at about 123 cm$^{-1}$ [19]. Depending on the degree of disorder (ratio of intermolecular to intramolecular interactions) this peak shifts continuously within the above wavenumber range.

Before proceeding to a more detailed discussion concerning the nature of the species composing the structure of a-GeTe we present in Fig. 3 a quantitative description of the vibrational modes with the aid of a Gaussian line fit procedure. The number of individual lines was the lowest that could provide a reasonable fit of the experimental data at 82, 300 and 373 K. Eight peaks were used; all of them have their distinct signature in the reduced Raman spectra. In total 24 parameters were optimized (peak areas, peak positions, and peak widths) resulting in the values tabulated in Table I. As mentioned above, bands B and C are composed of at least two peaks each, denoted as $B_1$, $B_2$ and $C_1$, $C_2$ on Fig. 3. Inspection of table I as well as Fig. 4 demonstrates that systematic changes do indeed take place upon heating the amorphous solid towards the crystallization temperature. Such changes (especially those in intensity) are not common in studies of other amorphous solids and thus deserve considerable attention. Summarizing, the analysis showed that: (i) peak positions monotonically shift to lower frequencies with increasing temperature in a moderate way as expected, and (ii) drastic changes in peak intensities take place implying an interplay between the species (structural units) participating in a-GeTe structure.

An obvious question arising from the above findings is why do so many vibrational modes emerge and what is their assignment in terms of structural units. The apparent dissimilarity between the Raman spectrum of a-GeTe and that of c-GeTe [cf. Fig. 1] rules out the adoption of a rocksalt-type structure or 3(Ge):3(Te) coordination in a-GeTe. Based on the tetrahedral coordination that Ge atoms favor and the twofold preferential coordination of Te atoms one expects the predominance of $GeTe_4$ tetrahedra. For chemically ordered structures, where homopolar Ge-Ge and Te-Te bonds are precluded, $GeTe_4$ tetrahedra are the only structural units for the stoichiometric composition $Ge_{0.33}Te_{0.67}$ where Ge and Te atoms are fourfold and twofold coordinated, respectively. However, the way of incorporation of extra Ge atoms into the structure, as the composition increases from 33 to 50%, is not obvious. Actually, many models have been advanced so far in the literature in order to tackle this problem. We describe below the most robust of these models discussing their predictions in relation to our results.

Models dealing with structural properties of $Ge_xCh_{1-x}$ (Ch: S, Se, Te) have frequently been focused on the Ch-rich binary systems due to the ease of glass formation and hence to the availability of experimental data (Raman and infrared) for direct comparisons. On the experimental side germanium sulfides [20] and selenides [21] have been studied in the glassy state (bulk glasses) using Raman spectroscopy for x up to 0.45 and 0.4, respectively, while amorphous thin films of germanium tellurides [10] have been studied over a much wider composition, for x up to 0.67. Among the models that have as a prerequisite the preservation of a fourfold coordination for Ge and a twofold coordination for Te, through covalent bonding, the most famous have been the chain-crossing model (CC) [7(b), 22], the random-covalent-network model (RCN) [7(b), 22], and the random bonding model [23].

The CC model presumes that Ge atoms act as crosslinks between the one dimensional chalcogens (Se, Te) chains. No more than 33% of Ge atoms can be accommodated in the structure. The basic ingredient of the model is the high degree of chemical order because at any composition x no Ge-Ge bonds are present. The model is valid in the chalcogen-rich phase ($0 < x < 0.33$) and hence it is not applicable in our case.

In the RCN model all types of bonds are allowed with equal probability, disregarding energetic factors (bond energies). The model is in principle applicable in the entire composition range and specifically predicts that the increase in the number of Ge-Te bonds with x is sublinear and the decrease in the number of Te-Te bonds with 1-x is superlinear.

The random bonding model proposed by Philipp [23] in an effort to account for optical properties of $SiO_x$ films. Being formulated for the concentration range $0.33 < x < 1$, this model is therefore operative for the germanium-rich regime. The model rules out the possibility of Ch-Ch bonds adopting a statistically weighted fraction of five types of tetrathedra of the type $GeCh_{4-n}Ge_n$ (n = 0, 1, 2, 3, 4). The distributions of these types of tetrahedral are shown in Fig. 5. At the stoichiometric composition x=0.33 the model predicts the existence of one type of tetrahedra, i.e. $GeTe_4$. This fraction is depleted rapidly with increasing x and at x=0.5, of interest here, the distribution consists of 6.25% $GeTe_4$, 25% $GeTe_3Ge$, 37.5 % $GeTe_2Ge_2$, 25% $GeTeGe_3$, and 6.25% $GeGe_4$.

Let us now proceed to the discussion of the theoretical predictions of tetrahedrally bonded solids. For a tetrahedral symmetry ($T_d$), group theory predicts four active Raman modes: the symmetric stretching mode $v_1(A_1)$, the bending modes $v_2(E)$, $v_4(F_2)$ located at frequencies lower than $v_1(A_1)$, and the antisymmetric stretching $v_3(F_2)$ mode at a frequency higher than $v_1(A_1)$ [24]. The irreducible representation of the Raman active modes in this case is $\Gamma(T_d) = A_1 + E + 2F_2$. Given the above context we can proceed to the following assignments of the Raman spectrum of a-GeTe. The two lowest Raman peaks at about 65 cm$^{-1}$ and 86 cm$^{-1}$ can undoubtedly be associated with $v_4(F_2)$ and $v_2(E)$ vibrational modes, respectively. The peak at 217 cm$^{-1}$ can be associated with the $v_3(F_2)$ mode. Therefore it remains to account for the bands located in the interval 100–200 cm$^{-1}$. To identify the stretching frequency $v_1(A_1)$ we can be based on the analogy between the Raman modes of the GeTe$_4$ and the GeI$_4$ tetrahedra [25]. Since the mass ratio of germanium tetraiodide is very similar to germanium tetratelluride we also expect a correspondence in the respective Raman frequencies. The frequencies of the isolated GeI$_4$ molecule are [26]: 60 cm$^{-1}$ [$v_2$(E)], 79 cm$^{-1}$ [$v_4$(F$_2$)], 159 cm$^{-1}$ [$v_1$(A$_1$)], and 264 cm$^{-1}$ [$v_3$(F$_2$)]. The ratio $v_3 / v_1 = 1.66$ in GeI$_4$ forces us to consider that the $v_1$(A$_1$) mode of the GeTe$_4$ tetrahedron is expected near 130 cm$^{-1}$; peak B$_2$ [cf. Fig. 4] is indeed located at this frequency. Following similar reasoning, Fisher *et. al.* [10] associated peak C (their poor resolution did not help resolving this peak in two components) with Te-Te bonds. As mentioned above, Te-Te vibrations in a-Te are found near 150 cm$^{-1}$; however, the existence of amorphous-like Te clusters (chains) in a-GeTe is highly unlikely and the assignment of peak C to Te-Te vibrations can safely be discarded. The adoption of the scenario that associates peak C with the $v_1$(A$_1$) mode and peak B with crystalline-like Te-Te vibrations [9] can also be considered implausible.

The arguments mentioned above force us to adopt the idea that the richness of the possible species envisaged in the random bonding model is probably reflected in the multiplicity of the vibrational lines illustrated in Fig. 3. Obviously, the $v_1$(A$_1$) vibrational frequency of the GeGe$_4$ tetrahedra is located at ~280 cm$^{-1}$ as mentioned above. Further, the vibrational modes of the Ge-rich tetrahedra, i.e. GeTeGe$_3$, are also expected not to differ considerably from those of the GeGe$_4$ species. Therefore, we are left with three tetrahedral species, namely GeTe$_4$, GeTe$_3$Ge, and GeTe$_2$Ge$_2$, with proportions ~6%, 25%, and 37.5 %, respectively. The first of these species is the heaviest one and hence its vibrational frequency is expected lowest. It is thus reasonable to associate peak B$_1$ at 111 cm$^{-1}$ with the $v_1$(A$_1$) vibrational frequency of the GeTe$_4$ tetrahedron which, however, accounts for the ~18% of the total reduced intensity in the spectral region 100–200 cm$^{-1}$. This is three times greater than the fraction expected from statistical consideration but it should be bear in mind at this point that the Raman intensity of a vibrational mode depends not only on the relative fraction of the species in question but also on the Raman coupling coefficient (cross section). This is related to the polarizability derivative of the vibrational mode, with respect to a generalized coordinate, during a vibrational period. Tellurium's high polarizability entails prominent Raman activity of the Te-rich tetrahedra. In addition to this, the higher symmetry of GeTe$_4$ (T$_d$) compared with C$_{3v}$ and C$_{2v}$ of the GeTe$_3$Ge and GeTe$_2$Ge$_2$ species, respectively, also accounts for the observability of the $v_1$(A$_1$) mode of GeTe$_4$.

The symmetric stretching modes of GeTe$_3$Ge and GeTe$_2$Ge$_2$ tetrahedra are expected at progressively higher wavenumbers than that of the GeTe$_4$ tetrahedra. The increased width of B$_2$ might imply that the spectral envelope of this line includes the modes of the aforementioned peaks. We prefer this assignment because the remaining two peaks $C_1$ and $C_2$ can be associated with either the extra Raman active modes due to symmetry lowering ($T_d \rightarrow C_{3v}$ and $T_d \rightarrow C_{2v}$) and/or the symmetric stretch modes of edge-sharing tetrahedra. Manifestations of edge-sharing tetrahedra in Raman spectra are common in tetrahedral glasses, see for example $ZnX_2$ (X: Cl, Br) [27] and $GeS_2$ [28]. In both cases, the frequency of the $v_1(A_1)$ vibrational mode associated with edge-sharing configurations is higher than that of the corresponding mode of apex-bridged species, which indicates species with stronger intra-tetrahedral bonding in the former. Interestingly, detailed composition dependence study of the Raman spectra in $Ge_xS_{1-x}$ and $Ge_xSe_{1-x}$ revealed that the $v_1(A_1)$ vibrational mode of the edge-sharing tetrahedra increases monotonically with increasing x, in the range 0<x<0.33 where experimental data are available [28]. Further, the effect in selenides is more prominent than sulfides, and hence we expect that it would even stronger for tellurides. This fact combined with the even higher value of x in our case (x=0.5) support the comparable intensity of modes B and C.

The validity of the assignments given above can be checked by considering the symmetry properties of the invoked tetrahedral species $GeTe_{4-n}Ge_n$ (n = 0, 1, 2). The depolarization ratio, $\rho$, of the Raman scattered signal, see Fig. 3(b) is indeed a sensitive indicator of the symmetry properties of vibrational modes. $\rho$ is defined as the ratio of the Raman intensity collected when the polarization of incident and scattered light are orthogonal over the Raman intensity collected when the polarization of incident and scattered light are parallel each other. The frequency of the minimum in $\rho$ (111 cm$^{-1}$) exhibits a striking coincidence with the position (112 cm$^{-1}$) of the $v_1(A_1)$ mode of the $GeTe_4$ tetrahedra, see Fig. 3. Obviously, the n=0 (or the $GeTe_4$) is the tetrahedron with the higher symmetry. The lowering of the symmetry of mixed tetrahedra, i.e. $C_{3v}$ for $GeTe_3Ge$ and $C_{2v}$ for $GeTe_2Ge_2$ implies a progressively higher value of $\rho$ as indeed Fig.3 reveals. Interestingly, out of the spectral region of the $v_1(A_1)$ modes of the various tetrahedra (100-140 cm$^{-1}$) $\rho$ acquires higher values indicating a less symmetric type of vibrational motion. The above findings are clear evidence in favor of the tetrahedra model in a-GeTe and in particular of the assignments we have provided to account for the Raman spectral details.

### 4.2. A structural mechanism for the a-GeTe → c-GeTe transition

Having formulated a rather plausible description of the structure of a-GeTe compatible with the experimental data we can now proceed to the elucidation of the structural transformations responsible for the amorphous-to-crystal transition. Based on the idea discussed above we propose a possible structural arrangement for the GeTe$_{4-n}$Ge$_n$ tetrahedra illustrated in Fig. 6. This particular arrangement shows apex- and edge-bridged tetrahedra where Te-Te bonds are absent. An important feature of this structural model is that with minor atomic rearrangements, including interchanges of few bonds, small displacement and/or tetrahedra rotations, the distorted rocksalt structure, in which c-GeTe crystallizes, can be obtained. Specifically, the atom denoted $Ge^c$ is surrounded by six Te atoms (encircled in blue color); three of them are closer to $Ge^c$ and the other three at a larger non-bonded distance. Analogously, the atom marked as $Te^c$ is also embedded in a local environment that offers octahedral coordination, namely, six Ge atoms encircled in red color. The relatively small structural changes needed for the atomic configuration shown in Fig. 6 in order to obtain the crystalline form might account for the rapidity of crystallization which is an important prerequisite in phase-change media such as 2GeTe – Sb$_2$Te$_3$ [2] where GeTe is the major component. The amorphous-to crystal transformation entails a mechanism which for a given type of atoms prevents atoms of the same kind to enter its first coordination shell. This implies that species where the tetrahedral positions are occupied by both types of atoms, Ge and Te, should become less favored with increasing temperature. Indeed, the population of the GeTe$_4$ increases with increasing temperature at the expense of other types of tetrahedra as can be seen from the temperature dependence of the B$_1$ / B$_2$ intensity ratio shown in Fig. 4. The formation of an increasing number of GeTe$_4$ will subsequently facilitate the transformation to the distorted GeTe$_6$ octahedra intrinsic to the c-GeTe structure.

The mechanism described here satisfies certain conditions that constitute the basis of optical cognitive information processing as proposed by Ovshinsky [29]. Indeed, the basic feature of devices that might be used for multistage data storage is the accessibility of partially amorphous states intermediate between the completely amorphous and the completely crystalline form. A prerequisite for this performance is a contrasting structural organization between the amorphous state and the crystal; a condition that is met in GeTe. Further, the diversity of the Ge-Te coordination number in various tetrahedra is indicative of the various stages of amorphicity. The closer to 0 the value of n of a local environment of GeTe$_{4-n}$Ge$_n$ (n = 0,1,2,3,4) tetrahedra, the smaller the structural (and hence the energy) difference between this region and the crystal. Therefore, the value of n can somehow be related to the degree of disorder thus corresponding to the different stages of local reflectivity or storage capability.

The unexpectedly different Raman spectrum of c-GeTe compared to that of a-GeTe can now easily be understood in terms of the aforementioned argumentation. Inspecting Fig. 1 we observe a severe softening in the frequency position of the Raman peaks. Vibrational band B exhibits a redshift of about 45 cm$^{-1}$; a value highly unusual for other amorphous-to-crystal transitions. Let us attempt a quantitative estimate of the expected redshift in the transition from the amorphous to the crystal. At first, the redshift is justified from the fact that the coordination number of Ge atoms changes from 4 to 6. This means that the electronic density of Ge is distributed to a larger number of bonds and hence the strength of the bonds is reduced. In addition, 3 of the Te atoms are situated at a larger distance which results in even weaker interactions with the central Ge atom. EXAFS studies of a-GeTe [8] have shown that the Ge-Te distance is of about 2.65 Å while in c-GeTe recent EXAFS data [9] result in 2.801 Å and 3.136 Å for short and long Ge-Te bonds, respectively.

The change in the coordination number induces relaxation in bond energy due to change in bond length. In our case where the amorphous-to-crystal transition induces bond dilation we have $d^{cr} = c\, d^{a}$, where $d^{cr}$ and $d^{a}$ is the bond length in the crystal and the amorphous state, respectively, and $c$ is a bond dilation coefficient with $c > 1$. The relation between the bond energy and the bond length is a power law dependence on the coefficient $c$ [30], i.e.

$$E(d^{cr}) = c^{-m} E(d^{a}) \qquad (2)$$

For a number of materials, experimental facts have shown that $m \approx 4$ [30]. Employing Eq. (2) and the bond lengths $d^{a} = 2.65$ Å, and $d^{cr} = 2.968$ (which is the mean value of the aforementioned values for c-GeTe) we obtain $c \approx 1.12$ and hence $E(d^{cr}) = 0.635\, E(d^{a})$. Considering the bond energies proportional to the bond frequencies we estimate using the above finding that the 80 cm$^{-1}$ peak of c-GeTe should appear at ~125 cm$^{-1}$. There is a surprising coincidence between this result and the frequency of the Raman band B of a-GeTe. This finding lends support to the assignment we have given to this band and to our suggestion that the involvement of Te-Te vibrations can be ruled out.

## 5. Conclusions

A temperature dependent Raman spectroscopic study of GeTe has been undertaken with the aid of a near IR laser in order to avoid thermo-structural and photo-structural changes. Emphasis was placed on the detailed structure of the amorphous form as well as on the amorphous-to-crystal transformation. Knowledge of the specific atomic rearrangements during this transformation is particularly important due to the fact that GeTe is the main component of the chalcogenide system (2GeTe-$Sb_2Te_2$) used as the active film in modern optical recording media such as DVDs. Employing a temperature reduction scheme and analyzing the complex Raman bands in individual vibrational modes made it possible to advance a detailed comparison between the data obtained at various temperatures. The analysis showed that the results are consistent with the random bonding model which (i) does not allow the existence of Te-Te bonds, and (ii) describes the structure as a combination of five types of tetrahedra, namely $GeTe_{4-n}Ge_n$ (n = 0, 1, 2, 3, 4). The identification of the peaks has been facilitated by a direct comparison of the normal modes of $GeI_4$. Taking into account the preference of tetrahedral glasses to form intermediate-range structural order with edge-bridged tetrahedra, we assigned particular Raman modes, which did not seem to fit with the tetrahedral symmetry modes of apex-bridged tetrahedra, to such edge-bridged species.

Using bond length – bond energy relations and recent EXAFS data we were able to explain the dramatic redshift of the Raman spectrum of crystalline GeTe with respect to that of the amorphous state. All the above facts corroborate the 4(Ge):2(Te) coordination in a-GeTe and further provide an atomic picture of the local structure compatible with the enhanced material's ability to transform from the amorphous to the crystalline state, which is the major advantage of materials utilized in phase-change optical data storage. Work in progress aims at elucidating the structure of the quasi-binary alloy GeTe-$Sb_2Te_3$.

Finally, we would like to state that the Raman band assignments given here could be considered as tentative and alternative interpretation may be proposed. However, the model proposed is compatible with the experimental observations as regards the intensities, the frequencies, as well as the polarization properties of the Raman bands. For example, the existence of $GeTe_6$ species with quasi-octahedral symmetry, analogous to those of the crystalline phase cannot be completely ruled-out. However, the strongest peak of c-GeTe is situated at ~80 cm$^{-1}$, that is, very close to the frequency region of the bending modes of the $GeTe_4$ tetrahedra, which makes the substantiation of octahedral species obscure.

## Acknowledgements

K. S. Andrikopoulos acknowledges financial support from the project "PYTHAGORAS I" funded by the Greek Ministry of National Education and Religion Affairs. S. N. Yannopoulos whishes to than Prof. G. N. Papatheodorou for illuminating discussions concerning structure of amorphous solids and Dr. A. Chrissanthopoulos for his help in producing Fig. 6.

**Table I:** Peak parameters (intensity and peak position) as determined through fits of the experimental data with a sum of eight Gaussian distributions. A possible assignment of these peaks originating from the GeTe$_{4-n}$Ge$_n$ tetrahedra is given in the first column.

| Peak identity [assignment] | 82 K | | 300 K | | 373 K | |
|---|---|---|---|---|---|---|
| | Intensity[a] (arb. units) | Frequency (cm$^{-1}$) | Intensity[a] (arb. units) | Frequency (cm$^{-1}$) | Intensity[a] (arb. units) | Frequency (cm$^{-1}$) |
| A$_1$ [ν$_4$(F$_2$)] | –[b] | –[b] | 235 | 65 | 98 | 63 |
| A$_2$ [ν$_2$(E)] | –[b] | –[b] | 266 | 88 | 139 | 86 |
| B$_1$ [ν$_1$(A$_1$), n=0, corner-sharing] | 137 | 112 | 236 | 111 | 288 | 109 |
| B$_2$ [ν$_1$(A$_1$), n=1,2, corner-sharing] | 528 | 128 | 456 | 127 | 368 | 125 |
| C$_1$ [ν$_1$(A$_1$) n=0, edge-sharing] | 295 | 165 | 508 | 162 | 516 | 160 |
| C$_2$ [ν$_1$(A$_1$) n=1,2 edge-sharing] | 130 | 183 | 142 | 182 | 101 | 180 |
| D [ν$_3$(F$_2$)] | 80 | 222 | 125 | 217 | 110 | 211 |
| E [c] | 74 | 150 | 44 | 145 | 186 | 141 |

[a] Spectra have not been normalized; thus comparison between individual peaks among various temperatures is meaningless.

[b] Strong elastic peak prohibited accurate determination.

[c] The assignment of peak E is uncertain; it was employed in the fitting procedure to increase fit quality.

**FIGURE CAPTIONS**

**Fig. 1:** Stokes-side raw Raman spectra of a-GeTe at 82, 300 and 373 K. The spectrum of c-GeTe at 300 K is also shown for comparison.

**Fig. 2:** Reduced Raman spectra of a-GeTe at 82, 300 and 373 K using Eq. 1. The reduced spectrum of c-GeTe at 300 K is also shown for comparison.

**Fig. 3:** (a) A representative example of the fitting procedure followed in order to unravel the fine structure of the reduced Raman spectra of a-GeTe at 300 K. (b) Depolarization ratio, $\rho = \dfrac{I_{HV}}{I_{VV}}$, where $I_{HV}$ is the spectrum obtained with the polarizations of the incident and scattered beams orthogonal, while $I_{VV}$ is the corresponding spectrum obtained when the two polarizations are parallel.

**Fig. 4:** Temperature dependence of the peak parameters extracted form the fitting procedure as described in the text. The main figure shows the normalized area ratio of the sub-peaks $B_1$, $B_2$, $C_1$, and $C_2$. Sub-peaks $B_1$ and $C_1$ increase by a factor of 3 at the expenses of their companions $B_2$ and $C_2$, respectively. The insets show the temperature dependence of the peak positions which show a continuous mild red-shift with increasing temperature.

**Fig. 5:** Relative distribution of the various types of $GeTe_nG_{1-n}$ tetrahedra, as a function of the chalcogens atomic fraction according to the Phillip's model of Ref. 23.

**Fig. 6:** (Color online.) A possible structural model of a-GeTe's short- and medium- range order. Small dark spheres and large gray spheres stand for Ge and Te atoms, respectively. The Te atoms encircled in blur color and the Ge atoms encircled in red color denote the atoms that will form the first coordination spheres of the dashed encircled Ge and Te atoms, respectively after the crystallization of the material.

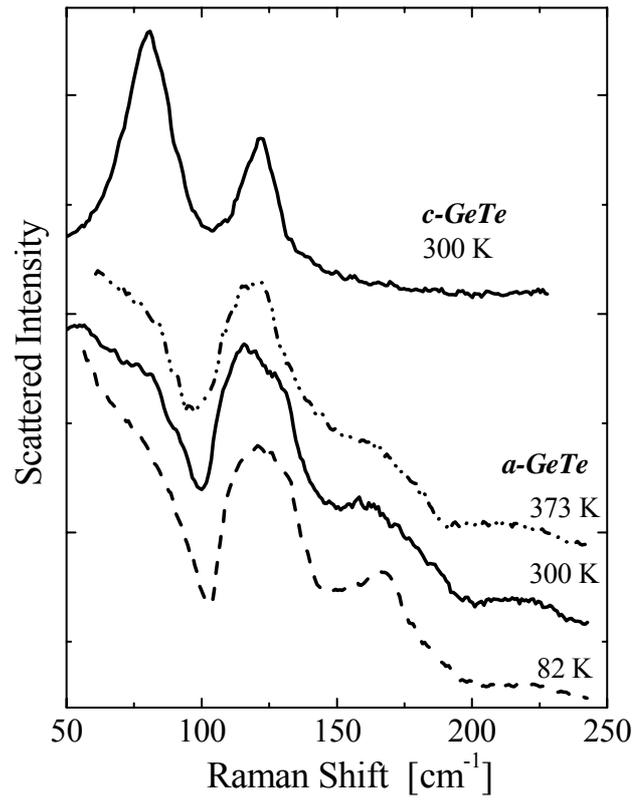

**Fig. 1**

Raman scattering study of the a-GeTe structure and possible mechanism for the amorphous-to-crystal transition

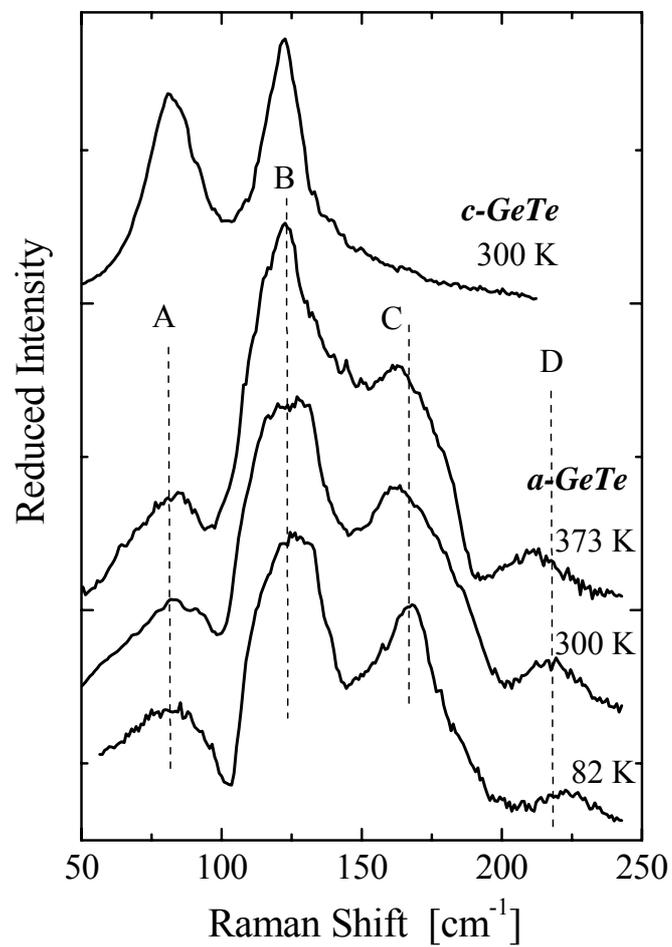

**Fig. 2**

Raman scattering study of the a-GeTe structure and possible mechanism for the amorphous-to-crystal transition

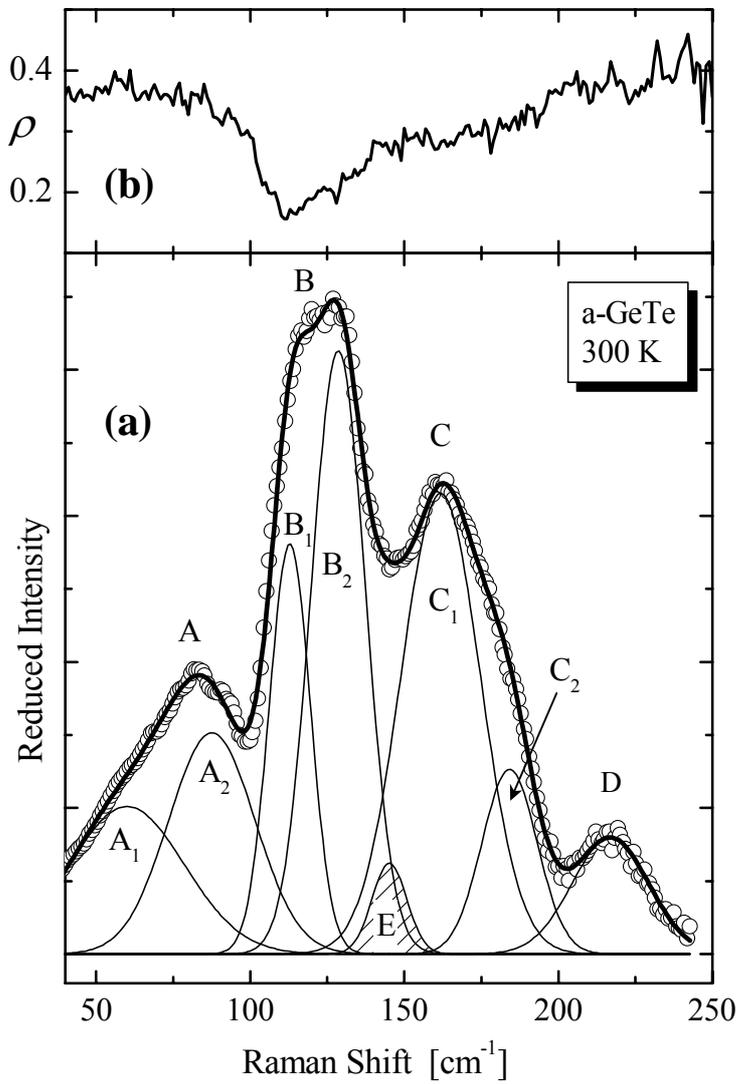

**Fig. 3**

Raman scattering study of the a-GeTe structure and possible mechanism for the amorphous-to-crystal transition

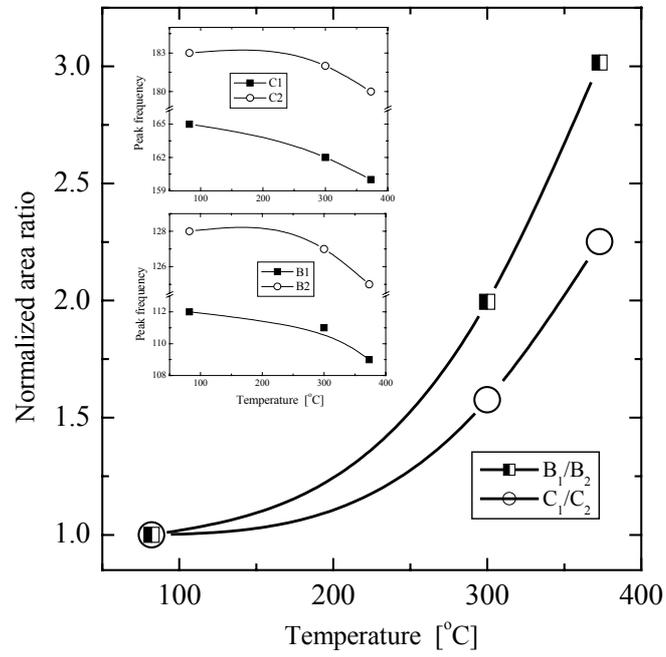

**Fig. 4**

Raman scattering study of the a-GeTe structure and possible mechanism for the amorphous-to-crystal transition

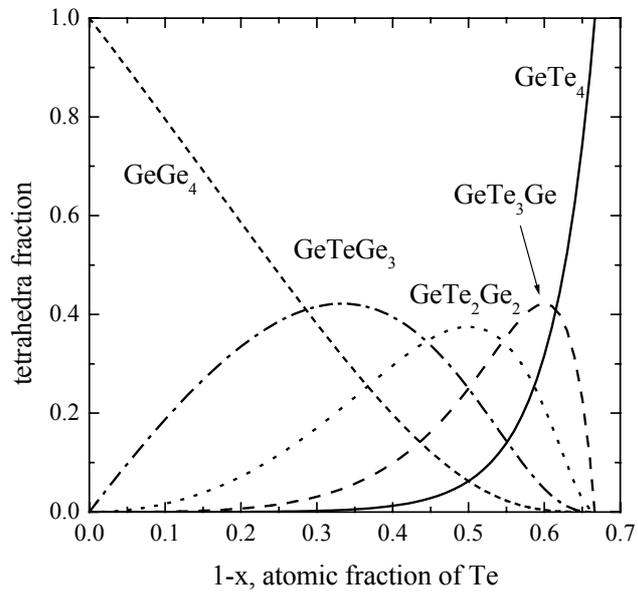

**Fig. 5**

Raman scattering study of the a-GeTe structure and possible mechanism for the amorphous-to-crystal transition

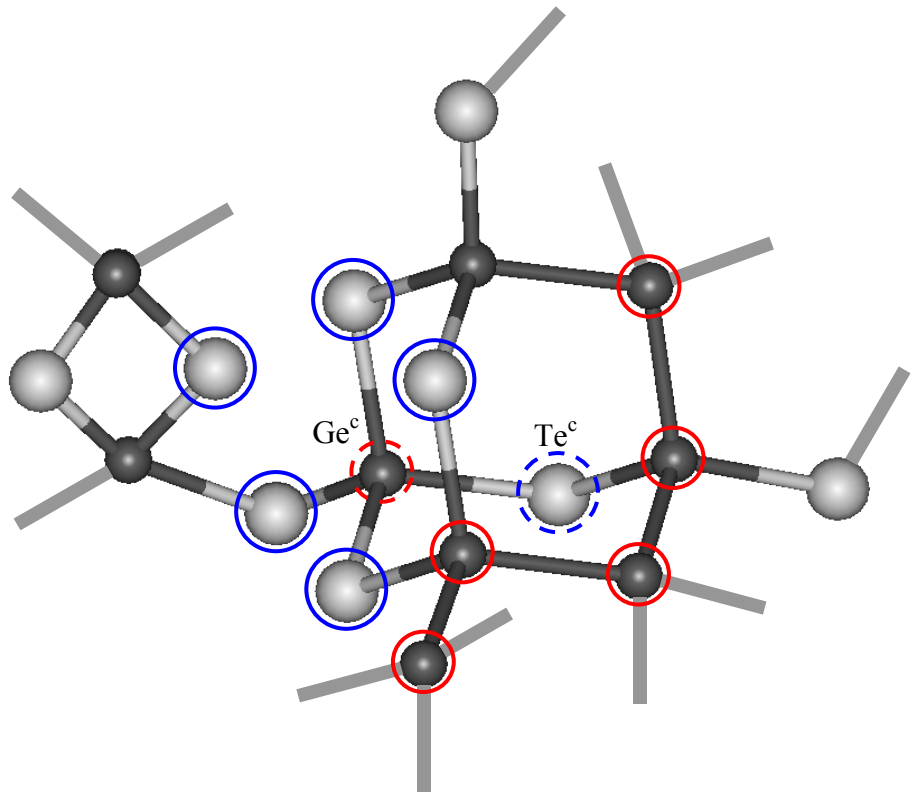

**Fig. 6**

(Color online) Raman scattering study of the a-GeTe structure and possible mechanism for the amorphous-to-crystal transition